\documentclass[twocolumn,aps,showpacs]{revtex4}
\usepackage{epsfig}
\begin{document}

\title{An upper limit on hypertriton production in collisions of Ar(1.76 AGeV)+KCl}

\author{G.~Agakishiev$^{7}$, D.~Belver$^{17}$, A.~Blanco$^{2}$, M.~B\"{o}hmer$^{10}$,
J.~L.~Boyard$^{15}$, P.~Cabanelas$^{17}$, E.~Castro$^{17}$, S.~Chernenko$^{7}$, M.~Destefanis$^{11}$,
F.~Dohrmann$^{6}$, A.~Dybczak$^{3}$, E.~Epple$^{9}$, L.~Fabbietti$^{9}$, O.~Fateev$^{7}$,
P.~Finocchiaro$^{1}$, P.~Fonte$^{2,b}$, J.~Friese$^{10}$, I.~Fr\"{o}hlich$^{8}$, T.~Galatyuk$^{5,c}$,
J.~A.~Garz\'{o}n$^{17}$, R.~Gernh\"{a}user$^{10}$, C.~Gilardi$^{11}$, M.~Golubeva$^{13}$, D.~Gonz\'{a}lez-D\'{\i}az$^{5}$,
F.~Guber$^{13}$, M.~Gumberidze$^{5,15}$, T.~Heinz$^{4}$, T.~Hennino$^{15}$, R.~Holzmann$^{4}$,
I.~Iori$^{12,e}$, A.~Ivashkin$^{13}$, M.~Jurkovic$^{10}$, B.~K\"{a}mpfer$^{6,d}$, T.~Karavicheva$^{13}$,
I.~Koenig$^{4}$, W.~Koenig$^{4}$, B.~W.~Kolb$^{4}$, R.~Kotte$^{6}$, A.~Kr\'{a}sa$^{16}$,
F.~Krizek$^{16}$, R.~Kr\"{u}cken$^{10}$, H.~Kuc$^{3,15}$, W.~K\"{u}hn$^{11}$, A.~Kugler$^{16}$,
A.~Kurepin$^{13}$, S.~Lang$^{4}$, J.~S.~Lange$^{11}$, K.~Lapidus$^{9}$, T.~Liu$^{15}$,
L.~Lopes$^{2}$, M.~Lorenz$^{8,c}$ $^{*}$, L.~Maier$^{10}$, A.~Mangiarotti$^{2}$, J.~Markert$^{8}$,
V.~Metag$^{11}$, B.~Michalska$^{3}$, J.~Michel$^{8}$, E.~Morini\`{e}re$^{15}$, J.~Mousa$^{14}$,
C.~M\"{u}ntz$^{8}$, L.~Naumann$^{6}$, Y.~C.~Pachmayer$^{8}$, M.~Palka$^{3}$, V.~Pechenov$^{4}$,
O.~Pechenova$^{8}$, J.~Pietraszko$^{4}$, W.~Przygoda$^{3}$, B.~Ramstein$^{15}$, L.~~Rehnisch$^{8}$,
A.~Reshetin$^{13}$, A.~Rustamov$^{8}$, A.~Sadovsky$^{13}$, P.~Salabura$^{3}$, T.~Scheib$^{8}$ $^{*}$,
A.~Schmah$^{a}$, H.~Schuldes$^{8}$, E.~Schwab$^{4}$, J.~Siebenson$^{9}$, Yu. G.~Sobolev$^{16}$,
S.~Spataro$^{f}$, B.~Spruck$^{11}$, H.~Str\"{o}bele$^{8}$, J.~Stroth$^{8,4}$, C.~Sturm$^{4}$,
A.~Tarantola$^{8}$, K.~Teilab$^{8}$, P.~Tlusty$^{16}$, M.~Traxler$^{4}$, R.~Trebacz$^{3}$,
H.~Tsertos$^{14}$, V.~Wagner$^{16}$, M.~Weber$^{10}$, C.~Wendisch$^{6,d}$, M.~Wisniowski$^{3}$,
J.~W\"{u}stenfeld$^{6}$, S.~Yurevich$^{4}$, Y.~Zanevsky$^{7}$}

\affiliation{
(HADES collaboration) \\\mbox{$^{1}$Istituto Nazionale di Fisica Nucleare - Laboratori Nazionali del Sud, 95125~Catania, Italy}\\
\mbox{$^{2}$LIP-Laborat\'{o}rio de Instrumenta\c{c}\~{a}o e F\'{\i}sica Experimental de Part\'{\i}culas , 3004-516~Coimbra, Portugal}\\
\mbox{$^{3}$Smoluchowski Institute of Physics, Jagiellonian University of Cracow, 30-059~Krak\'{o}w, Poland}\\
\mbox{$^{4}$GSI Helmholtzzentrum f\"{u}r Schwerionenforschung GmbH, 64291~Darmstadt, Germany}\\
\mbox{$^{5}$Technische Universit\"{a}t Darmstadt, 64289~Darmstadt, Germany}\\
\mbox{$^{6}$Institut f\"{u}r Strahlenphysik, Helmholtz-Zentrum Dresden-Rossendorf, 01314~Dresden, Germany}\\
\mbox{$^{7}$Joint Institute of Nuclear Research, 141980~Dubna, Russia}\\
\mbox{$^{8}$Institut f\"{u}r Kernphysik, Goethe-Universit\"{a}t, 60438 ~Frankfurt, Germany}\\
\mbox{$^{9}$Excellence Cluster 'Origin and Structure of the Universe' , 85748~Garching, Germany}\\
\mbox{$^{10}$Physik Department E12, Technische Universit\"{a}t M\"{u}nchen, 85748~Garching, Germany}\\
\mbox{$^{11}$II.Physikalisches Institut, Justus Liebig Universit\"{a}t Giessen, 35392~Giessen, Germany}\\
\mbox{$^{12}$Istituto Nazionale di Fisica Nucleare, Sezione di Milano, 20133~Milano, Italy}\\
\mbox{$^{13}$Institute for Nuclear Research, Russian Academy of Science, 117312~Moscow, Russia}\\
\mbox{$^{14}$Department of Physics, University of Cyprus, 1678~Nicosia, Cyprus}\\
\mbox{$^{15}$Institut de Physique Nucl\'{e}aire (UMR 8608), CNRS/IN2P3 - Universit\'{e} Paris Sud, F-91406~Orsay Cedex, France}\\
\mbox{$^{16}$Nuclear Physics Institute, Academy of Sciences of Czech Republic, 25068~Rez, Czech Republic}\\
\mbox{$^{17}$LabCAF. F. F\'{\i}sica, Univ. de Santiago de Compostela, 15706~Santiago de Compostela, Spain}\\
\\
\mbox{$^{a}$ now at Lawrence Berkeley National Laboratory, ~Berkeley, USA}\\
\mbox{$^{b}$ also at ISEC Coimbra, ~Coimbra, Portugal}\\
\mbox{$^{c}$ also at ExtreMe Matter Institute EMMI, 64291~Darmstadt, Germany}\\
\mbox{$^{d}$ also at Technische Universit\"{a}t Dresden, 01062~Dresden, Germany}\\
\mbox{$^{e}$ also at Dipartimento di Fisica, Universit\`{a} di Milano, 20133~Milano, Italy}\\
\mbox{$^{f}$ now at Dipartimento di Fisica Generale and INFN, Universit\`{a} di Torino, 10125~Torino, Italy}\\
\mbox{$^{\ast}$ corresponding authors: m.lorenz@gsi.de, t.scheib@gsi.de}\\
}

\begin{abstract}
A high-statistic data sample of Ar(1.76 AGeV)+KCl events recorded with HADES is used to search for a hypertriton signal. An upper production limit per centrality-triggered event of $1.04$ x $10^{-3}$ on the 3$\sigma$ level is derived. Comparing this value with the number of successfully reconstructed $\Lambda$ hyperons allows to determine an upper limit on the ratio $N_{_{\Lambda}^3H}/N_{\Lambda}$, which is confronted with statistical and coalescence-type model calculations. 

\end{abstract}


\maketitle
The hyperon-nucleon (Y-N) interaction plays an important role in nuclear physics. For example, the appearance of $\Sigma N$ cusp structures in elementary reactions \cite{HIRES,COSY} or the formation of quasi-molecules like the $\Lambda(1405)$ \cite{Lambda1405} can be related to peculiarities of the YN interaction. Even our understanding of fundamental astrophysical objects like neutron stars demands a precise knowledge about this interaction: Depending on the strength of the YN interaction, the core of neutron stars might consist either of hyperons, strange quarks or a state with a kaon condensate, thus constraining the upper limit of neutron star masses \cite{neutron_stars}. The spectroscopy of hypernuclei, i.e.\ nuclei containing one or more hyperons, provides a tool for studying details of the YN interaction, which is partly responsible for their binding and hence their life-time \cite{K1,K2,cong,K3}.

Recently, in addition to results from reactions induced by cosmic rays \cite{Cosmic}, secondary meson \cite{Pions, Mesons, Mesons2, Mesons3} or primary electron beams \cite{electrons}, also data from relativistic heavy-ion collisions have been published. Most prominently is the discovery of the anti-hypertriton by the STAR collaboration at a center of mass energy of $\sqrt{s_{NN}}=$200 GeV \cite{RHIC}. But also at the lower beam energies signals have been reconstructed \cite{ags} and experimental programs are pursued. For instance in the kinetic beam energy range of 1-2 AGeV as provided by the SIS18 accelerator at the GSI Helmholtzzentrum for heavy-ion research in Germany, the HypHI collaboration recently published life time data for the hypertriton \cite{HypHi,HypHi1}. In contrast to elementary particle beams, heavy-ion collisions are not limited to hypernuclei close to the valley of $\beta$ stability and can therefore provide access to a larger variety of these states. The production cross section is an essential measurement in context of a comprehensive survey of strangeness degrees of freedom in heavy-ion collisions and future programs at FAIR \cite{CBM}. 

The formation of composite particles is an interesting issue in heavy-ion collisions. While at low and intermediate energies describing the formation of fragments is a challenging but meanwhile well understood task \cite{Aichelin1, Aichelin2, Dorso, Nield, Abdura}, at relativistic energies two descriptions are usually employed. In one model fragment formation is addressed via the coalescence of hadrons emerging from the collision zone \cite{Nagle, Scheibl}, while the other assumes a thermal population of all hadronic and nucleonic states \cite{pbm}. In the transition region, where the kinetic beam energy of nucleons is comparable with their rest mass, one needs more data, in particular for rare clusters with strangeness, to distinguish between the two approaches. The thermal model populates degrees of freedom (hadrons and their resonances as well as clusters) according to statistical weights. Coalescence formation of hypernuclei results due to the final state interaction between the involved baryonic objects and can in principle proceed along several paths.  For example one can speculate about sequential multistep processes leading to $_{\Lambda}^3$H, "formation at once" in a three-baryon clusterization or strangeness exchange on preformed $^3$H. The first step towards understanding the origin of strange hypernuclei in heavy-ion collisions at intermediate energies is to provide a constraint on the multiplicities. Here we focus on the hypertriton, i.e.\ $_{\Lambda}^3$H and provide an upper limit for its formation in the reaction Ar(1.76 AGeV)+KCl.

In this study we used a high-statistic data set of about $7.37\times10^{8}$ events recorded with the High-Acceptance Di-Electron Spectrometer (HADES) to search for the production of hypertritons.
HADES \cite{HHades} is a charged-particle detector consisting of a 6-coil toroidal magnet centered around the beam axis and six identical detection sections covering polar angles between $18^{\circ}$ and $85^{\circ}$. Each sector is equipped with a Ring-Imaging Cherenkov (RICH) detector followed by Multi-Wire Drift Chambers (MDCs), two in front of and two behind the magnetic field, as well as a scintillator wall. Hadron identification is based on the measurement of its momentum, its time-of-flight and on its energy-loss information in the scintillators, as well as in the MDC tracking chambers. 

The data presented here were acquired with an argon beam of $\sim 10^6$ particles/s at a beam energy of 1.76 AGeV incident on a four-fold segmented KCl target with a total thickness corresponding to $3.3$ $\%$ interaction probability. A fast diamond start detector located upstream of the target intersected the beam and delivered the time-zero information. The data readout was started by a first-level trigger (LVL1) requiring a charged-particle multiplicity MULT$ \ge16$ in the scintillator detectors. Based on a GEANT \cite{GEANT} simulation including the detector response and using Ar+KCl events generated with the UrQMD transport model \cite{UrQMD}, we found that the event ensemble selected by this (LVL1) trigger condition has a mean number of participating nucleons ($\langle A_{part}\rangle)$ equal to $38.5 \pm 3.9$. 

\begin{figure}[tb]
  \centering
\mbox{\epsfig{figure=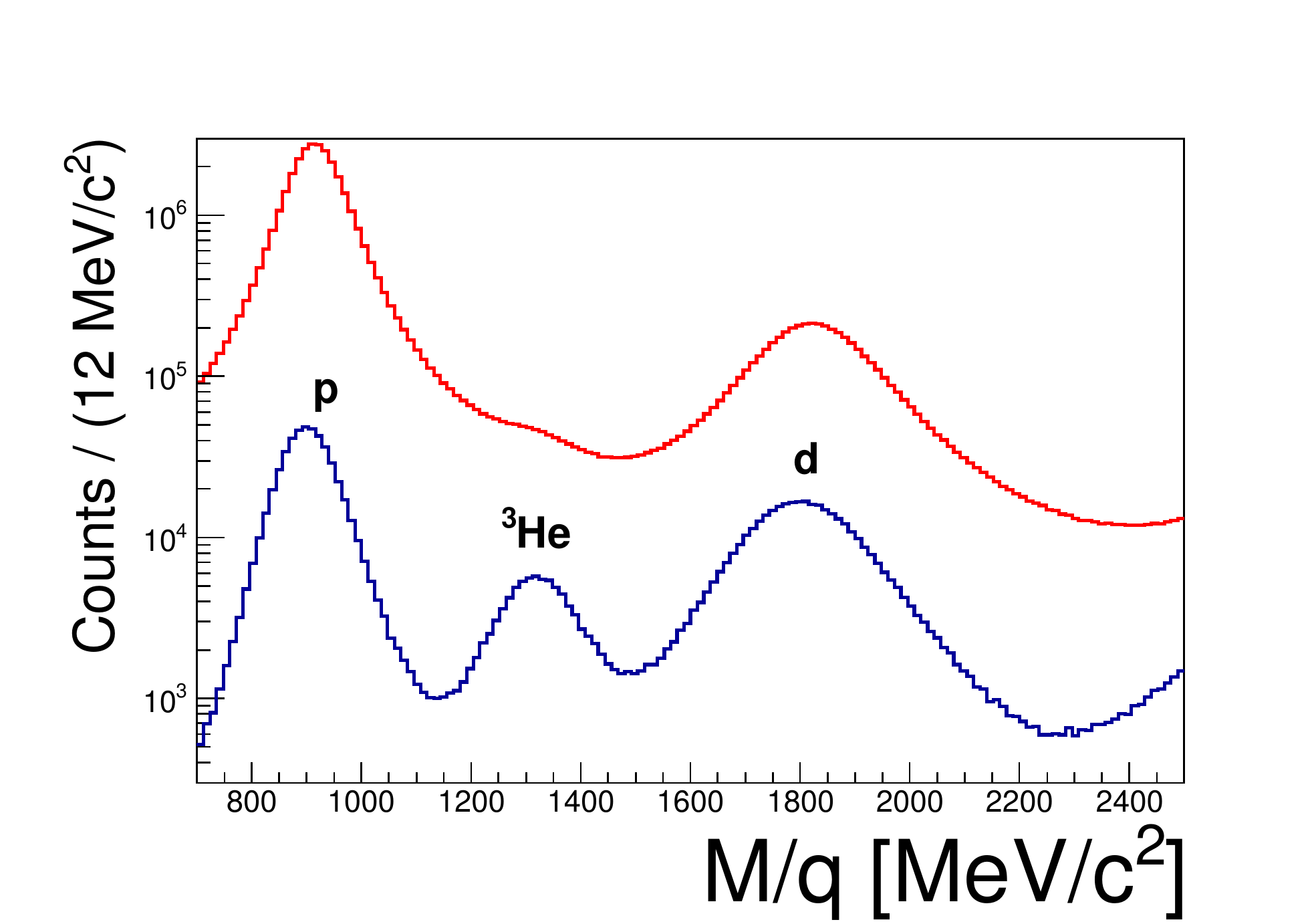, width=1.0\linewidth}}
\caption{Distribution of the mass over charge ratio as derived from the time-of-flight measurement before (upper curve) and after (lower curve) application of cuts on the energy-loss of $^3$He and on additional track quality criteria.}
  \label{He3}
\end{figure}
The average life-time of hypertritons is expected to be comparable to the one of the $\Lambda$ hyperon ($c\tau_{\Lambda}=7.9$ cm). 
Hypertritons $_{\Lambda}^3$H were searched for via the decay $_{\Lambda}^3$H $\rightarrow \pi^{-} + ^{3}$He. The branching ratio for this decay is about $25\%$ \cite{K2,cong}.
Hence, an important issue of the analysis was an identification of the two final-state particles with high purity. For details on the pion identification cf. \cite{schmah}. The selection of $^3$He ions is shown in Fig.~\ref{He3}. Cuts are applied on the quality of the tracks and the energy-loss. The shift of the $^3$He peak position can be fully reproduced by the GEANT simulation and is mainly due to the large energy-loss in the tracking detectors.

For further background suppression, various constraints on the decay topology, such as distance between the primary vertex and the decay vertex ($d_{V0}>20$ mm), minimal distance between the $\pi^-$ trajectory and the primary vertex ($d_{\pi^-}>10$ mm), distance of closest approach between the two tracks of the decay particles ($d_{dca}<8$ mm) and distance of the hypertriton track to the primary vertex ($d_{_{\Lambda}^3H}<6$ mm) 
were applied. To improve the decay vertex resolution a lower limit was required for the opening angle ($\alpha_{\hspace{1mm} ^{3}He,\pi^{-} } > 17^{\circ}$) between the two daughter particles. 
The cuts on the decay topology were optimized using hypertriton decays simulated with PLUTO \cite{Pluto} and processed with the same GEANT simulation. 
The hypertriton mass was set to $m_{_{\Lambda} ^3H} = 2991$ MeV/c$^2$. This value has been precisely measured by various experiments \cite{K3, keyes}.
The rapidity coverage of HADES for $_{\Lambda}^3$H relative to the center-of-mass rapidity of 0.86 ranges from -0.7 to 0.1.
In general one would expect the hypertriton rapidity distribution to be a folding of rapidity distributions from particles carrying strangeness and heavier clusters. In order to estimate the maximum effect of the limited acceptance, two extreme cases have been used: A single gaussian shaped rapidity distribution in accordance with a single static Maxwell-Boltzmann source located at mid-rapidity as observed for kaons \cite{HHades, schmah},     
or a rapidity distribution having a pronounced two-peak structure as observed for light fragments in this collision system \cite{Heidi}. In the second case the peaks are located at $y=\pm0.59$ in the center of mass system having a width of $\sigma=0.35$. The value for the slope parameter of the transverse momentum distributions $T_{eff}=100$ MeV is taken in both cases as the one that was found for the $\Lambda$ hyperon in the same collision system \cite{HHyp}. 
The width of the rapidity distribution from a single static thermal source located at mid-rapidity is then approximately (for $T_{eff} << {m_{^{3}_{\Lambda}H}}$) given by \[\sigma_{y} = \sqrt{\frac{T_{eff}}{m_{^{3}_{\Lambda}H}c^{2}}}.\] 
To get a realistic estimate of the uncorrelated background, the simulated signal was embedded into real events before track reconstruction. 


The fraction of detectable hypertritons due to the acceptance $\epsilon_{acc}$ and the reconstruction efficiency $\epsilon_{rec}$ of the hypertriton were determined in simulations to $\epsilon_{acc} \cdot \epsilon_{rec} (twopeak)  =0.03\%$ for those generated using a two-peak rapidity distribution and $\epsilon_{acc} \cdot \epsilon_{rec} (singlepeak) =0.043\%$ for those generated using a single-peak rapidity distribution. The acceptance $\epsilon_{acc}$ includes a branching ratio of $25\%$ \cite{K2,cong}.



\begin{figure}[tb]
  \centering
\mbox{\epsfig{figure=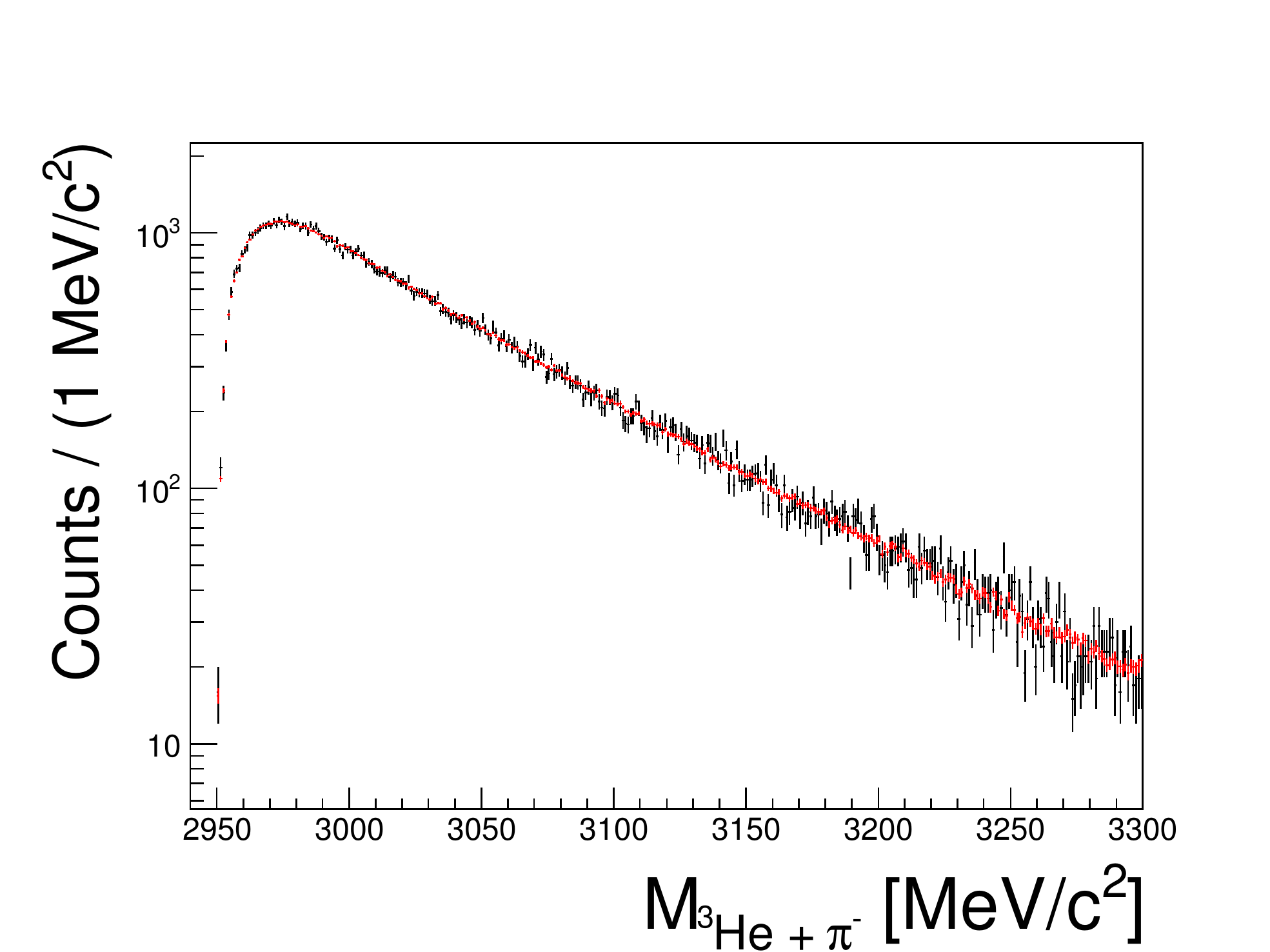}, width=1.0\linewidth}}
\mbox{\epsfig{figure=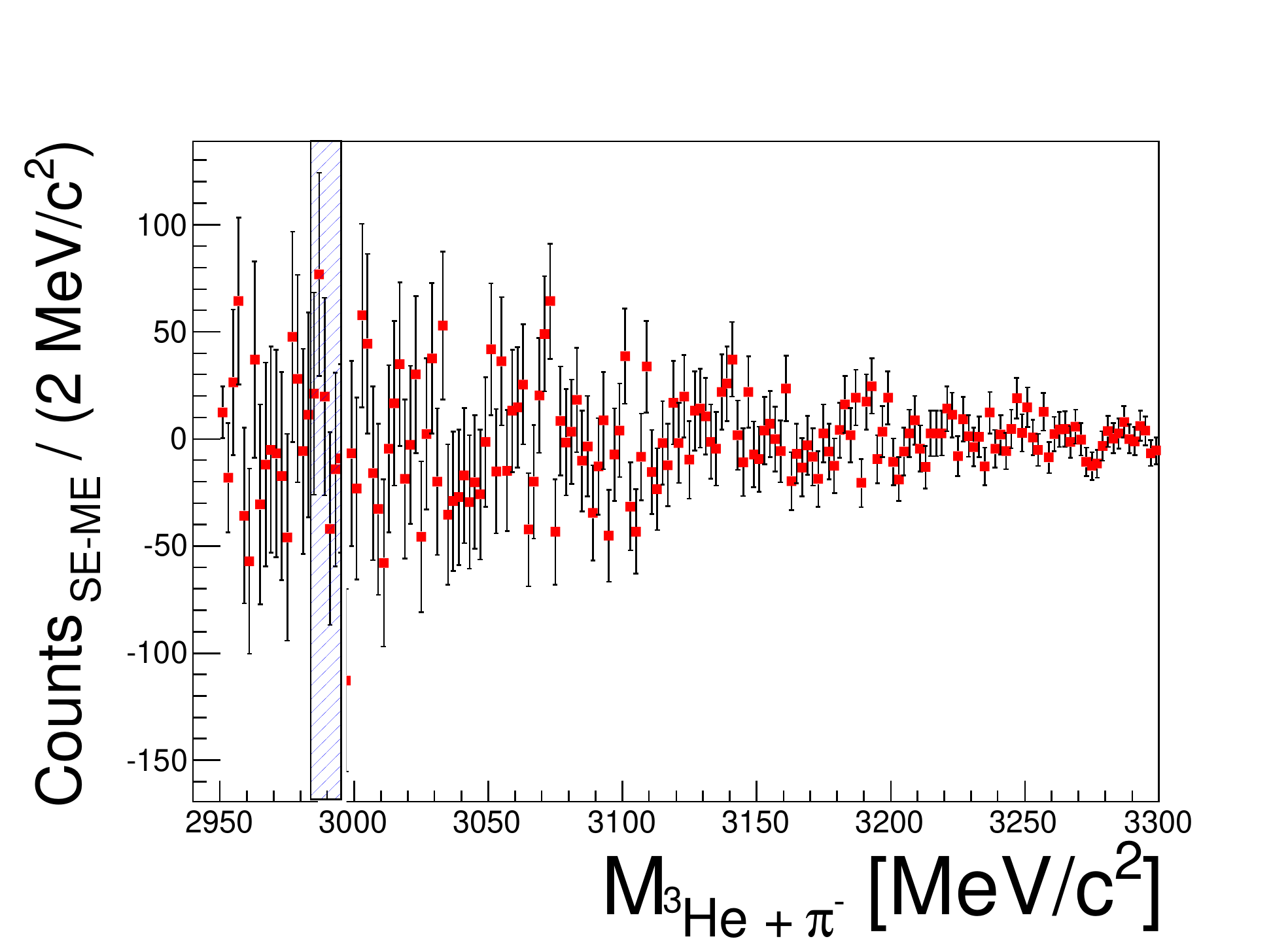}, width=1.0\linewidth}}
\caption{Top: Invariant-mass spectrum of $^3$He and $\pi^-$ exhibiting the pair combinations within one event ("same event") and within different events ("mixed event"). Both distributions are normalized to one another within the range 3010-3250 MeV/c$^{2}$. Bottom: Invariant mass spectrum of $^3$He and $\pi^-$ after subtraction of the mixed-event background from the experimental data. The hatched region depicts the invariant masses where a $_{\Lambda}^3$H signal is expected.}
  \label{Hypertriton2}
\end{figure}

In Fig.~\ref{Hypertriton2} (top), the invariant-mass distribution of the real data as well as the mixed-event background are shown. The background is obtained by combining $\pi^-$-$^3$He pairs coming from different events mixed within the same classes of multiplicity and target segments. Fig.~\ref{Hypertriton2} (bottom) illustrates that no significant hypertriton signal can be identified after background subtraction. Also a variation of the normalization interval (3010-3250 MeV/c$^2$) does not unravel an indication for a peak-like structure at the expected $_{\Lambda}^3$H mass (hatched region).

\begin{figure}[htb]
  \centering
\mbox{\epsfig{figure=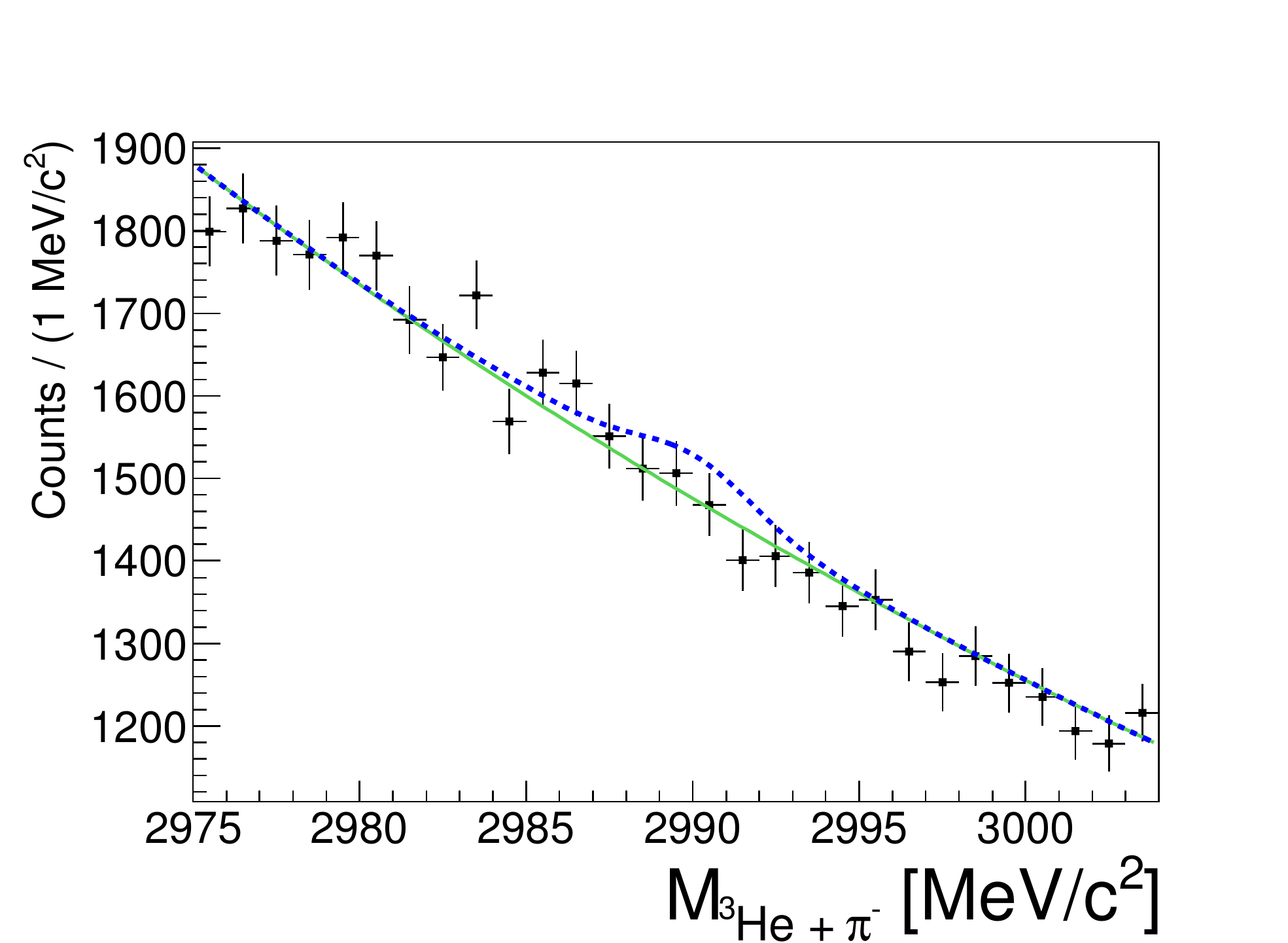}, width=1.0\linewidth}}
\caption{Invariant-mass spectrum of $^3$He and $\pi^-$ zoomed into the expected hypertriton region. The blue dashed line indicates the hypertriton signal according to simulations within the coalescence model assuming the upper limit to be the production multiplicity, superimposed to an exponential fit to the background (green continuos line).}
  \label{Hypertriton3}
\end{figure}

However, with this null result, an upper production limit can be derived. For this purpose the Feldman-Cousins approach \cite{feldman} is used, which is based upon the statistics of available data. For this case the presented invariant-mass distribution and the mixed-event background of Fig.~\ref{Hypertriton2} (top) are integrated in a $\pm3\sigma_{Gauss}$ region (where the width $\sigma_{Gauss}=2.5 $ MeV/c$^2$ is estimated from GEANT simulations) around the expected mean value of the signal ($m=2991$ MeV/c$^{2}$). With a chosen confidence level of $99.7\% \hspace{2mm}(\hat{=}\hspace{1mm} 3 \sigma)$ the method provides a multiplicity of $M_{FC}(twopeak)=(7.04 \pm ^{1.31}_{0.79}  \pm ^{2.04}_{1.64})$ x $10^{-4}$ and $M_{FC}(singlepeak)=(4.88 \pm 0.92  \pm ^{1.45}_{1.12})$ x $10^{-4}$ respectively.
The first error is calculated by varying the above mentioned topology cut values by 10$\%$ of the absolute value, while the second one is obtained by modifying the integration region of the hypertriton mass around the expected peak value (from $\pm1$ to $\pm5$ $\sigma_{Gauss}$). As the first limit $M_{FC}(twopeak)$ lies above the second, it is chosen as the upper limit on the hypertriton production.
Therefore, the positive errors are added to the extracted multiplicity which finally leads to the limiting value of $M_{UL}=1.04$ x $10^{-3}$ per (LVL1) triggered event. Fig.~\ref{Hypertriton3} shows again the invariant-mass spectrum of real data zoomed into the expected hypertriton signal region. In addition, the hypertriton peak is indicated assuming the calculated upper limit to be the actual multiplicity. The width is taken from simulated hypertritons which have passed a full GEANT simulation.



As an additional cross check of the method, simulated hypertritons were embedded into real data at such a ratio that a reconstruction is possible on a significance level comparable to the one obtained using the Feldman-Cousins approach for the upper limit. The corresponding multiplicity agrees within errors with our upper limit. 
The upper production limit fits well into the systematics on sensitivity of the apparatus obtained for other rare hadronic probes decaying off vertex in the Ar+KCl system like e.g.\ the $\Xi^{-}$ baryon \cite{HHyp,HPhi,HXi}.

Considering the upper limit, an upper boundary on the ratio of the rate of hypertritons to reconstructed $\Lambda$ hyperons \cite{HHyp} was calculated $N_{_{\Lambda}^3H}/N_{\Lambda}<(2.5 \pm 0.3)$ x $10^{-2}$. Note that the reconstructed $\Lambda$ hyperon signal includes the feed down from the slightly heavier $\Sigma^{0}$ baryon which decays into a $\Lambda$ and a photon with nearly $100\%$ branching ratio. Therefore, the $\Lambda$ multiplicity includes the contribution from $\Sigma^{0}$ decays.

The calculated ratio $N_{_{\Lambda}^3H}/N_{\Lambda}$ can be confronted with theoretical model calculations.
In general, the production rates of light hypernuclei relative to $\Lambda$ hyperons develop a pronounced maximum in the FAIR energy domain ($E_{kin}\approx 3-7$ AGeV) as a consequence of the transition from a baryon dominated to a meson dominated system \cite{pbm,steinheimer1}. In the statistical hadronization model, this behavior can be understood in terms of a strongly increasing baryon chemical potential $\mu_{B}$ and a decreasing temperature $T$. In addition, the yields decrease again towards lower energies due to local strangeness conservation and hence canonical suppression. The authors of \cite{pbm} predict a ratio $N_{^3_{\Lambda}H}/N_{\Lambda}(thermal)\approx 1$ x $10^{-2}$ for central heavy-ion collisions at our beam energy using their realization of the above mentioned statistical model.  
This value is in fair agreement with the value of $N_{^3_{\Lambda}H}/N_{\Lambda}(hydro)= 1.6$ x $10^{-2}$ extracted from a hydro calculation using the SHASTA algorithm \cite{steinheimer1,steinheimer2,73}.
On the other hand, cascade transport codes plus a dynamical coalescence model used for clusterization tend to predict lower values like $N_{^3_{\Lambda}H}/N_{\Lambda}(coalescence)\approx 1$ x $10^{-3}$ in \cite{zhangsong}  
using the ART 1.0 cascade code \cite{27}. While a combination of the Dubna Cascade Model plus a different coalescence model and the Quark-Gluon String Model \cite{67,63} give a value of $N_{^3_{\Lambda}H}/N_{\Lambda}(coalescence)= 8.3$ x $10^{-3}$ in \cite{steinheimer1,steinheimer2}. The first value for the coalescence production is for a system of comparable size (Ar+Ar), and the difference in beam energy for the predicted ratio is negligible according to the authors \cite{zhangsong2}.
The second value is again predicted for central collisions of heavy nuclei.
Our value of $N_{_{\Lambda}^3H}/N_{\Lambda}<(2.5 \pm 0.3)$ x $10^{-2}$ is very close to the various predicted ratios. However, to distinguish between the different production mechanisms more precise data and calculations for the exact experimental conditions like system size and centrality selections are needed.  
  

In summary, we used a high-statistic data sample of Ar(1.76 AGeV)+KCl collisions recorded with HADES to search for hypertriton production. With a confidence level of $3\sigma$ we provide an upper production limit of $M_{UL}=1.04$ x $10^{-3}$ per (LVL1) triggered event for the hypertriton, which agrees with the expectation considering the sensitivity of the experimental setup as determined by other rare strange particles such as the $\Xi^{-}$. From the upper limit the ratio of hypertritons to the successfully reconstructed $\Lambda$ hyperons could be calculated to be $<(2.5 \pm 0.3)$ x $10^{-2}$ and confronted with model calculations. 
Our limiting value for hypertriton production relative to the $\Lambda$  is very close to the predicted ratio by various models. However, to distinguish between the different production mechanisms implemented in the models, more precise data are needed.


We thank Y. Leifels, N. Herrmann, Y. Zhang and J. Steinheimer for fruitful discussions and gratefully acknowledge the support by 
LIP Coimbra, Coimbra (Portugal) PTDC/FIS/113339/2009, SIP JUC Cracow, Cracow (Poland) N N202 286038 NN202198639, FZ Dresden-Rossendorf (FZD), Dresden (Germany) BMBF 05P09CRFTE, 05P12CRGHE, TU M\"{u}nchen, Garching (Germany) MLL M\"{u}nchen DFG EClust 153 VH-NG-330 BMBF 06MT9156 TP5, GSI TMKrue 1012, NPI AS CR, Rez (Czech Republic) GACR 13-067595 and AS CR M100481202, USC - S. de Compostela, Santiago de Compostela (Spain) CPAN:CSD2007-00042, Goethe-University, Frankfurt (Germany) HA216/EMMI HIC for FAIR (LOEWE) BMBF:06FY9100I GSI F$\&$E, TU Darmstadt (Germany): VH-NG-823, Helmholtz Alliance HA216/EMMI, IN2P3-CNRS (Orsay). 

%
%
%


{}

\end{document}